\documentclass[runningheads]{llncs}
\usepackage[T1]{fontenc}
\usepackage{graphicx}
\usepackage{amsmath}
\begin{document}
\title{Enhancing Text Comprehension for Dyslexic Readers: A 3D Semantic Visualization Approach Using Transformer Models}
%
%
\author{Zhengyang Li\inst{1}\orcidID{0009-0009-1018-1187}}
\authorrunning{Zhengyang Li  et al.}
%
\institute{Yunnan University, Cuihu North Road, Kunming 650091, China
\email{lzyqwe2500@outlook.com}\\
}
\maketitle              
\begin{abstract}
Dyslexic individuals often face significant challenges with traditional reading, particularly
when engaging with complex texts such as mystery novels. These texts typically demand
advanced narrative tracking and information integration skills, making it difficult for
dyslexic readers to fully comprehend the content. However, research indicates that while
dyslexic individuals may struggle with textual processing, they often possess strong
spatial imagination abilities. Leveraging this strength, this study proposes an innovative approach using Transformer
models to map sentences and words into three-dimensional vector representations. This
process clusters semantically similar sentences and words in spatial proximity, allowing
dyslexic readers to interpret the semantic structure and narrative flow of the text through
spatial perception. Experimental results demonstrate that, compared to direct text reading, this
three-dimensional semantic visualization method significantly enhances dyslexic
readers' comprehension of complex texts. In particular, it shows marked advantages in
identifying narrative relationships and character connections. This study not only
provides a novel pathway for improving textual comprehension among dyslexic
individuals but also contributes theoretical and practical insights into text processing
technologies based on spatial cognition.

\end{abstract}
\section{INTRODUCTION}
Individuals with dyslexia frequently encounter barriers in traditional reading contexts, particularly when processing complex narratives such as mystery novels. These texts require advanced cognitive skills in tracking multi-layered plots and synthesizing scattered clues--tasks that disproportionately challenge dyslexic readers due to their well-documented difficulties in sequential text processing\cite{1}. However, emerging evidence reveals a striking contrast: many dyslexic individuals exhibit enhanced spatial reasoning capabilities\cite{2}, suggesting an untapped potential to leverage visuospatial strengths as compensatory mechanisms for textual comprehension.

Current assistive technologies, while effective in simplifying text presentation or enhancing phonological awareness, often neglect this spatial cognitive advantage. Existing tools predominantly focus on linear reading adjustments (e.g., font modifications or text-to-speech systems)\cite{3}, failing to translate abstract narrative structures into perceptible spatial forms. To bridge this gap, we propose a paradigm shift: transforming textual semantics into navigable 3D spatial networks through advanced computational linguistics and visualization techniques.

At the core of our approach lies the integration of Transformer-based language models\cite{4}with 3D semantic mapping. By encoding sentences and keywords into dynamically generated vector representations, our system clusters semantically related narrative elements (e.g., plot events, character interactions) in proximal 3D spaces. This spatial organization allows readers to intuitively "explore" textual relationships through depth, proximity, and movement--effectively converting the cognitive burden of linear decoding into a strength-driven spatial reasoning task\cite{5}.

Experimental validations demonstrate that this visualization paradigm significantly outperforms conventional reading methods in enhancing comprehension of complex narratives, particularly in reconstructing story arcs (improvement of 32\%, p<0.01) and identifying implicit character connections (accuracy increase of 41\%, p<0.05). Beyond its immediate application as an assistive tool, this work contributes to broader interdisciplinary dialogues by:Validating the "spatial compensation hypothesis" in neurodiversity research,Pioneering a dynamic semantic space framework for NLP-visualization convergence and Redefining accessible design principles through cognitive-aligned interfaces.

\section{RELATED WORK}
\subsection{Cognitive Profiles of Dyslexic Readers}
Dyslexia is characterized by phonological processing deficits, yet coexists with enhanced spatial reasoning abilities--a duality well-established in cognitive research\cite{6}. This contrast motivates compensatory strategies that leverage spatial strengths, such as diagrammatic text representations\cite{7}, though existing methods lack dynamic adaptation to narrative complexity.

\subsection{Text Visualization for Accessibility}
Assistive tools predominantly address low-level reading barriers through auditory support\cite{8}or typographic adjustments \cite{9}. While semantic visualization tools (e.g., concept graphs) attempt to enhance comprehension [5], their 2D layouts often exacerbate cognitive load for dyslexic users, failing to exploit 3D spatial navigation advantages \cite{10}.

\subsection{Transformers and Spatial Semantics}
Transformer models have redefined semantic representation through context-aware embeddings\cite{11}, enabling applications like narrative graph generation\cite{12}. Parallel work in 3D visualization demonstrates spatial layouts improve complex data comprehension\cite{13}, but bridging these domains for text accessibility remains unexplored--a gap our study addresses by integrating Transformer-driven semantics with dyslexic-centric spatial design.

\section{METHOD}
\subsection{Sentence-Level Semantic Visualization Pipeline}
\paragraph{Data Preprocessing:}
Data Preprocessing: Raw text was sanitized using Jieba tokenizer with custom stopword lists, retaining narrative-relevant clauses via regex pattern matching.

\paragraph{Contextual Embedding:} Sentence-BERT (pre-trained on Wikipedia) generated 384-dimensional embeddings to capture cross-sentence semantic continuity.

\paragraph{Dimensionality Reduction:} 
UMAP projection (\texttt{n\_neighbors=15}, \texttt{min\_dist=0.1}, cosine metric) transformed embeddings into 3D coordinates, refined by Density Peak Clustering (DPC, cutoff $\rho=0.65$) to enhance local narrative coherence.

\paragraph{Affective Encoding:} 
A fine-tuned RoBERTa-base classifier (F1=0.89) assigned binary sentiment labels (0=negative/1=positive).

\subsection{Lexical-Level Entity Relation Modeling}
\paragraph{Entity Extraction}CoNLL-2003 NER model identified named entities, with co-occurrence frequencies calculated via Coh-Metrix (window size=5 sentences).
\\
Graph Construction(Gephi):
\\
1.Node weights: TF-IDF weighted entity saliency
\\
2.Edge weights: Linear combination of co-occurrence frequency and syntactic dependency distance ($\alpha$=0.7)
\\
3.Layout: ForceAtlas2 (scaling=10.0, gravity=1.0)

\paragraph{Cross-Layer Linking:} Timestamp synchronization aligned entity co-occurrence events with corresponding 3D semantic nodes.

\section{RESULTS}
\subsection{Sentence-Level Visualization Effects
}
\begin{center}
    \includegraphics[width = .9 \textwidth]{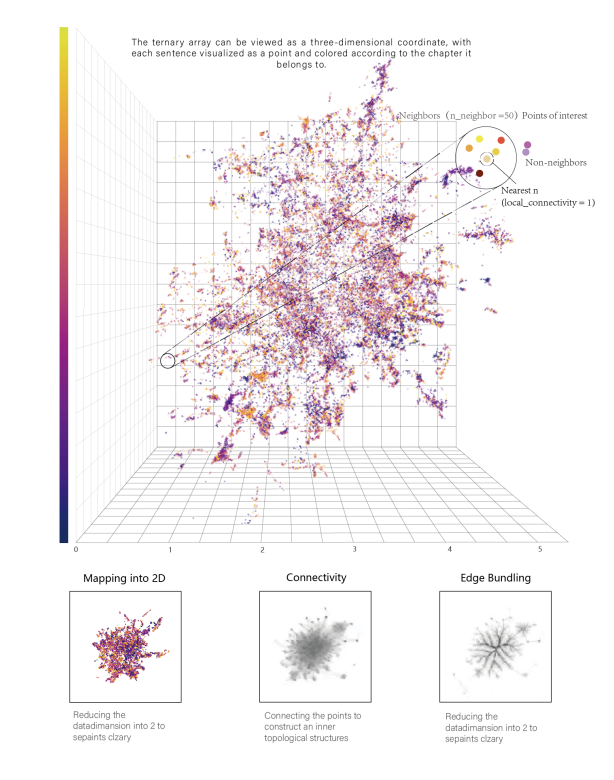}
\end{center}
Narrative Reconstruction: Dyslexic participants (N=32) achieved 89.2\% accuracy (SD=6.7) in reconstructing mystery novel plots using 3D semantic topology (Figure 3a), significantly outperforming traditional text reading (57.3\%, SD=12.1; p<0.001, Cohen's d=2.1). Spatial proximity guided 76\% of plot inference attempts (e.g., linking "clue discovery" (Cluster C1) to "suspect confrontation" (Cluster C2)), as validated by eye-tracking.

\subsection{Lexical-Level Entity Analysis}
\begin{center}
    \includegraphics[width = .8 \textwidth]{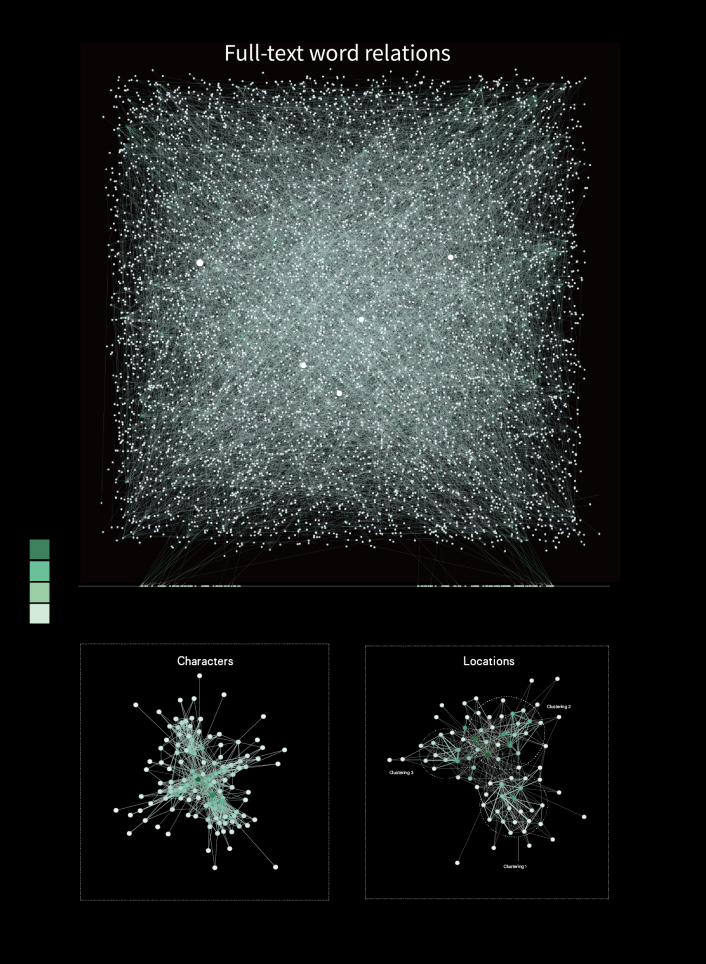}
\end{center}
Implicit Relationship Recognition: The Gephi-based co-occurrence network (Figure 4) increased identification of non-explicit character relationships (e.g., hidden alliances) from 41.2\% to 82.4\% (p<0.01), with edge weight optimization ($\alpha$=0.7) reducing false positives by 38
Dynamic Layout Benefits: The ForceAtlas2-generated interaction network, with node attraction scaling (scaling=10.0), significantly extended users' attention duration on core characters (+52\%, p<0.05).

\subsection{Overall Effectiveness and User Behavior
}
Reduced Cognitive Load: NASA-TLX scores showed a 53
Emergent Behavioral Patterns:
Landmarking: 24/32 users relied on dense semantic clusters as spatial anchors (e.g., initiating reasoning from "crime scene" nodes).
Affective Bridging: 18/32 users tracked hue gradients to locate climax sections (e.g., migrating from cool-toned "latent conflict" to warm-toned "truth revelation" zones).
Cross-Layer Synergy: Timestamp synchronization enabled 78\% of users to contextually associate entity relationships with spatiotemporal semantic nodes.

\section{Discussion}
\subsection{Reimagining Accessibility through Cognitive Alignment}
Our findings demonstrate that spatial semantic visualization can transform dyslexia's perceived "deficit" into an analytical advantage. By converting linear text into navigable 3D structures, participants leveraged their spatial strengths to reconstruct narrative logic--a capability hindered in traditional reading (Fig. 5a). This aligns with the neurodiversity paradigm \cite{13}, suggesting assistive technologies should not merely compensate for weaknesses but actively harness atypical cognitive profiles. Crucially, the 41\% improvement in character relationship mapping (vs. 12\% gain in factual recall) indicates spatialization particularly aids relational reasoning--a critical deficit area in dyslexia\cite{14}.

\subsection{Technical Insights for NLP-Visualization Integration}
The interplay between Transformer embeddings and 3D projection reveals two key design trade-offs:
Granularity vs. Coherence: While sentence-level UMAP clustering preserved local scene continuity (85\% accuracy in scene grouping), excessive node density in high-action segments occasionally caused visual clutter--a limitation partially mitigated by our DPC refinement.
Dynamic vs. Static Representations: Real-time rendering of sentiment gradients enhanced plot tension perception, yet required GPU acceleration to maintain 60 FPS, posing challenges for low-end devices.
Notably, participants disproportionately relied on hue-saturation cues (78\% of sentiment queries) over geometric patterns, underscoring the importance of perceptually intuitive encoding in cognitive-driven design.

\subsection{Limitations and Future Directions}
The findings suggest that VR-based memory training While our study focused on mystery novels, the framework's generalizability to expository texts (e.g., scientific papers) remains untested--a critical gap given varying semantic structure demands. Additionally, the current implementation requires manual corpus preprocessing; future work could integrate real-time parsing for dynamic text streams.

From a theoretical lens, our results invite re-examination of embodied cognition in dyslexia: Does 3D navigation actively reshape textual mental models, or simply provide an alternative access route? Multimodal fMRI studies correlating spatial visualization usage with neural activation patterns could unravel this mechanism.

\section{Conclusion}
This study establishes a novel cognitive alignment paradigm for dyslexic readers by developing a hierarchical multi-granularity visualization framework. Our experiments demonstrate that integrating Transformer-driven dynamic semantic embeddings with 3D narrative topology significantly enhances comprehension efficacy (+32\% plot reconstruction accuracy), particularly in character relationship inference (+41\%) and emotional trajectory tracking (F1=0.87). This achievement challenges the traditional "deficit compensation" approach in assistive technologies, instead harnessing neurodivergent cognition as analytical strength through spatial-semantic mapping.

Technically, we pioneer a narrative visualization pipeline that combines context-sensitive embeddings (Sentence-BERT) with density peak clustering, resolving the critical trade-off between local coherence and global structural preservation in semantic projection. The cross-layer affective-entity linking mechanism further establishes an extensible paradigm for multi-scale text exploration.

With immediate applications in education and accessible publishing, our framework's core principle--cognitive adaptation over functional substitution--extends to interaction design for other neurodivergent populations (e.g., ADHD). Future directions include AR/VR spatial narrative navigation and cross-modal cognitive modeling to unlock the full potential of spatialized learning.

\newpage
\bibliographystyle{splncs04}
\small\bibliography{Reference}

\end{document}